# An Advanced Epitaxial Strategy Enabling Vertical GaN Devices on Silicon Wafers

Fumio KAWAMURA*, Takeyoshi ONUMA, Kazutaka MITSUISHI*

Corresponding Authors: Fumio Kawamura, and Kazutaka Mitsuishi

**Keywords:** gallium nitride, silicon wafers, heteroepitaxy, buffer layers, vertical power devices

[1]Research Center for Electronic and Optical Materials, National Institute for Materials Science (NIMS),

1-1 Namiki, Tsukuba, Ibaraki 305-0044, Japan

Phone: +81-29-860-4428

E-mail: KAWAMURA.Fumio@nims.go.jp

[2]Department of Applied Physics, School of Advanced Engineering and Department of Electrical

Engineering and Electronics, Graduate School of Engineering, Kogakuin University, 2665-1 Nakano,

Hachioji, Tokyo 192-0015, Japan

Phone: +81-42-628-4704

E-mail: onuma@cc.kogakuin.ac.jp

[3]Center for Basic Research on Materials, National Institute for Materials Science (NIMS), 1-2-1 Sengen,

Tsukuba, Ibaraki 305-0047, Japan

Phone: +29-863-5474

E-mail: MITSUISHI.Kazutaka@nims.go.jp

**Abstract**

While vertical gallium nitride (GaN)-on-silicon architectures promise a transformative leap in cost-effective power electronics and high-resolution micro-LEDs, their deployment remains bottlenecked by the high electrical resistance of conventional epitaxial buffer layers. Here, a universal and straightforward sputtering-based strategy is presented to realize high-quality GaN epitaxial films on Si(111) substrates characterized by exceptionally low vertical resistance, ohmic behavior, and robust thermal stability. This technique centers on the *in-situ* formation of a sub-nanometer (~0.5 nm) silicide-based template via rapid thermal annealing—a method demonstrating unprecedented versatility across 25 different metallic species. Scanning transmission electron microscopy (STEM) reveals that a unique amorphous-like interlayer (AL-IL) effectively accommodates lattice mismatch and relaxes epitaxial strain. These AL-IL templates further serve as high-performance platforms for metalorganic chemical vapor deposition (MOCVD) overgrowth, successfully bridging the gap between scalable, low-cost fabrication and device-grade vertical performance.

## 1. Introduction

The development of vertical Gallium Nitride (GaN) devices is anticipated to dramatically enhance the performance of next-generation electronics, from power transistors to light-emitting diodes (LEDs), and is currently a subject of intensive global research [1–10]

In power electronics, vertical GaN transistors are poised to offer superior efficiency and higher power-handling capabilities, surpassing even the widely adopted Silicon Carbide (SiC) based counterparts. Similarly, in the burgeoning field of GaN-based micro-LEDs for high-resolution, low-electric consumption displays, a vertical architecture enabling backside electrode formation is projected to slash fabrication costs, facilitate further miniaturization, and improve brightness uniformity. However, a critical barrier to the widespread adoption of these vertical devices is the prohibitive cost of native GaN single-crystal substrates. While growth methods such as hydride vapor phase epitaxy (HVPE) [11–15], the ammonothermal method [16–21], and the Na-flux method [22–29] can produce such substrates, their high expense remains a fundamental obstacle. Moreover, the intrinsic cost of the raw material, gallium, places a fundamental limit on cost reduction. Conversely, realizing vertical GaN devices on silicon wafers offers a compelling path toward drastic cost reduction. Recent advances, such as the insertion of an AlGaN/GaN superlattice, have enabled the commercialization of cost-effective GaN-on-Si lateral transistors [30–39]. The high electrical resistance of this superlattice buffer, however, precludes vertical current flow, confining device architectures to lateral configurations [40]. Therefore, the development of a buffer layer that facilitates the epitaxial growth of high-quality GaN on silicon while maintaining low vertical resistance represents the critical breakthrough needed to unlock the potential of vertical GaN-on-Si devices [41–46].

## 2. Results and Discussion

### 2.1 Versatile ultrathin buffer strategy

We have developed a novel, low-resistance buffer layer that enables the heteroepitaxial growth of high-crystallinity GaN films on silicon substrates.

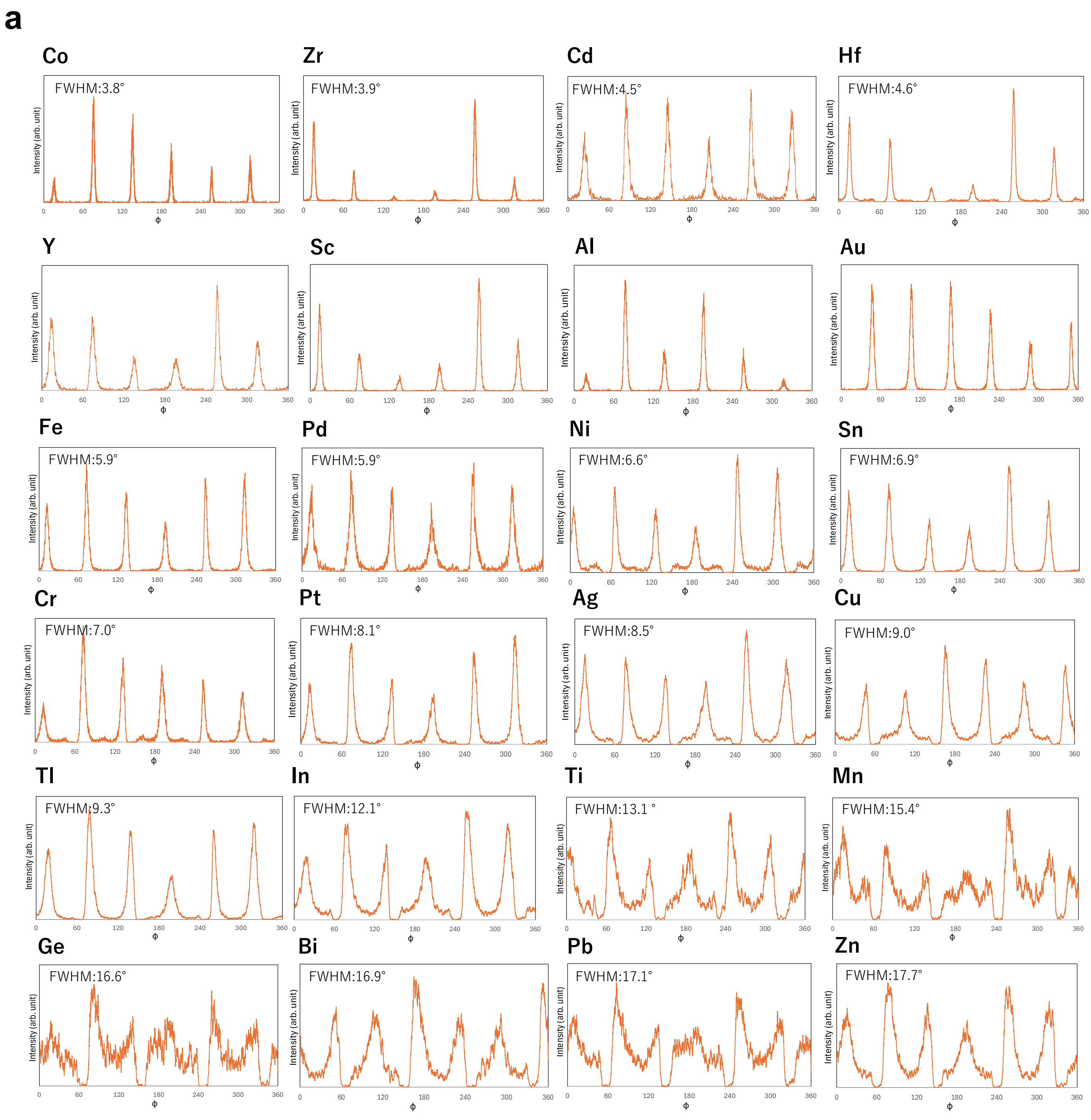

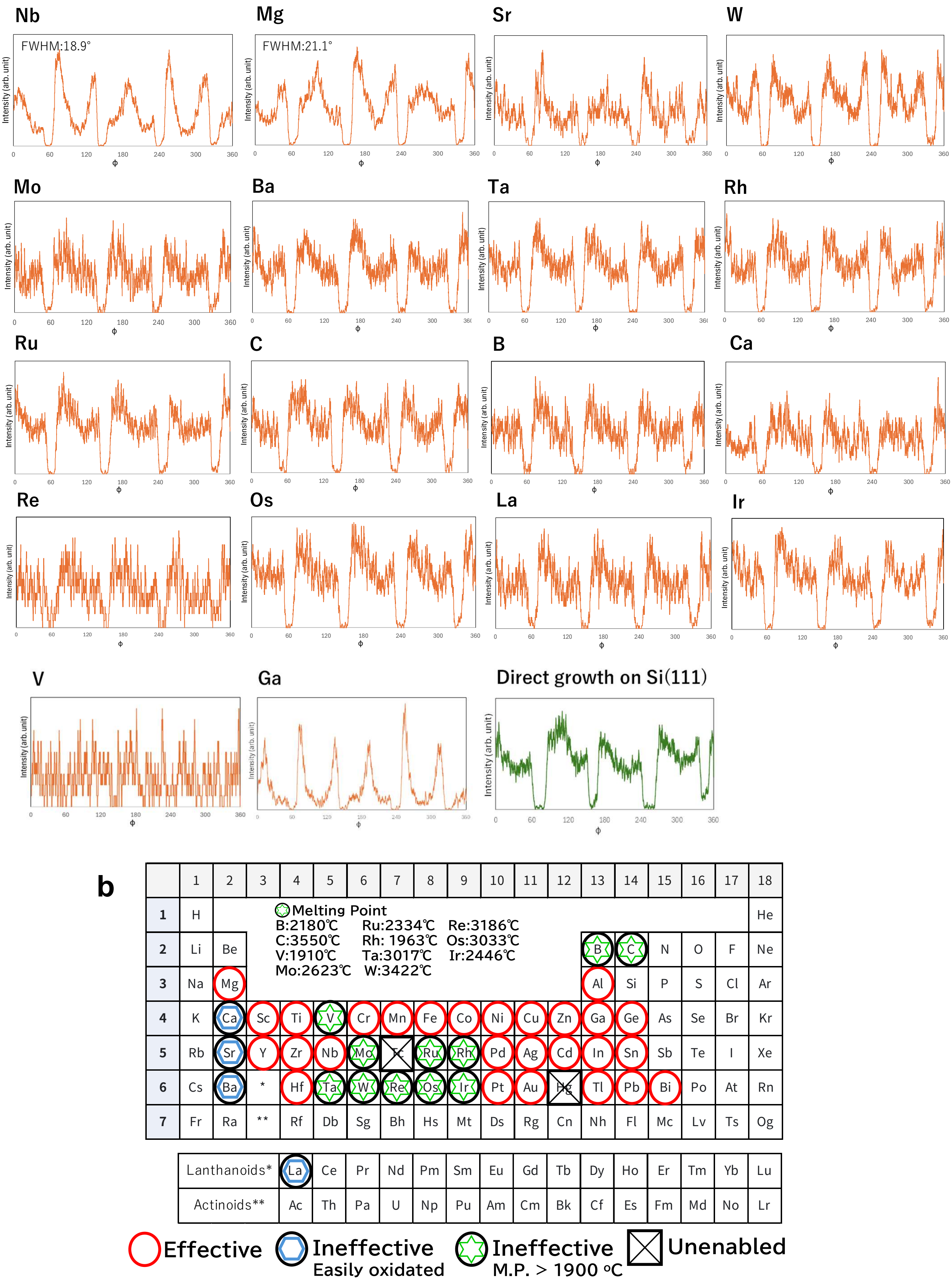
Nb
FWHM:18.9°
Mg
FWHM:21.1°
Sr
W
Mo
Ba
Ta
Rh
Ru
C
B
Ca
Re
Os
La
Ir
V
Ga
Direct growth on Si(111)
Intensity (arb. unit)
ϕ
b
Melting Point
B:2180℃ Ru:2334℃ Re:3186℃
C:3550℃ Rh: 1963℃ Os:3033℃
V:1910℃ Ta:3017℃ Ir:2446℃
Mo:2623℃ W:3422℃
Lanthanoids*
Actinoids**
Effective
Ineffective
Easily oxidated
Ineffective
M.P. > 1900 ºC
Unenabled

**Figure 1. Crystallographic quality of sputtered GaN films as a function of the pre-sputtered element.** a) The data are ordered from best to worst epitaxial alignment. The identity of the pre-sputtered element for each film is noted in the upper left. In cases where six-fold symmetric peaks are observed, the full width at half-maximum (FWHM) of the strongest peak is denoted in each profile, quantifying the degree of in-plane orientation. As an exception for the case of Ga deposition, a buffer layer was formed by rapidly heating the GaN film grown on Si(111) up to 1200 °C to thermally decompose the GaN and induce a reaction between the resulting Ga and the Si substrate. b) Based on the results in Fig. 1a, samples exhibiting sixfold symmetric GaN diffraction peaks were classified as 'effective', while others were categorized as 'ineffective'.

**Fig. 1a** demonstrates the pivotal role of an ultrathin buffer layer in enabling the epitaxial growth of sputtered GaN films on Si(111) substrates. Note that all experiments were conducted on 3°-offcut Si(111) substrates. The epitaxial quality of GaN grown using such ultrathin buffer layers is sensitive to factors including substrate off-cut angle, annealing temperature, and impurity levels (**Supporting Information Figs. 1-4**).

To identify an optimal buffer material, we conducted a systematic screening of 43 different elements. Each element was deposited as a 0.5-nm-thick interlayer, annealed at 725°C, and subsequently capped with a 400-nm-thick GaN film. The success of GaN epitaxy, defined by the observation of six-fold rotational symmetry in its diffraction pattern (**Fig. 1b**), reveals a distinct trend. The initial growth mode of GaN persists even as the film thickness reaches 1 μm, indicating a stable epitaxial process throughout the deposition (**Supporting Information Fig. 5**). We find that elements ineffective in promoting epitaxy fall into two categories: 1) refractory metals with melting points above 1900°C, and 2) highly oxophilic elements. Conversely, all other tested elements successfully enabled epitaxial growth. This indicates that a 0.5-nm-thick film of nearly any element with a melting point below 1900°C and moderate oxidation resistance can form a viable template for GaN epitaxy upon high-temperature annealing. We propose that the failures of the epitaxial growth are due to two distinct mechanisms: the refractory metals do not react with silicon to form a silicide interlayer, while the highly oxophilic elements are prematurely oxidized by residual gases in the chamber, which prevents silicidation. These results strongly suggest that the formation of an ultrathin silicide interlayer—irrespective of its specific composition—is the key factor enabling the subsequent GaN epitaxy.

## 2.2 Mechanism of epitaxy via amorphous-like interlayer

To validate our hypothesis that the silicide interlayer is the key enabler, we investigated the effect of annealing temperature using a 0.5-nm cobalt film.

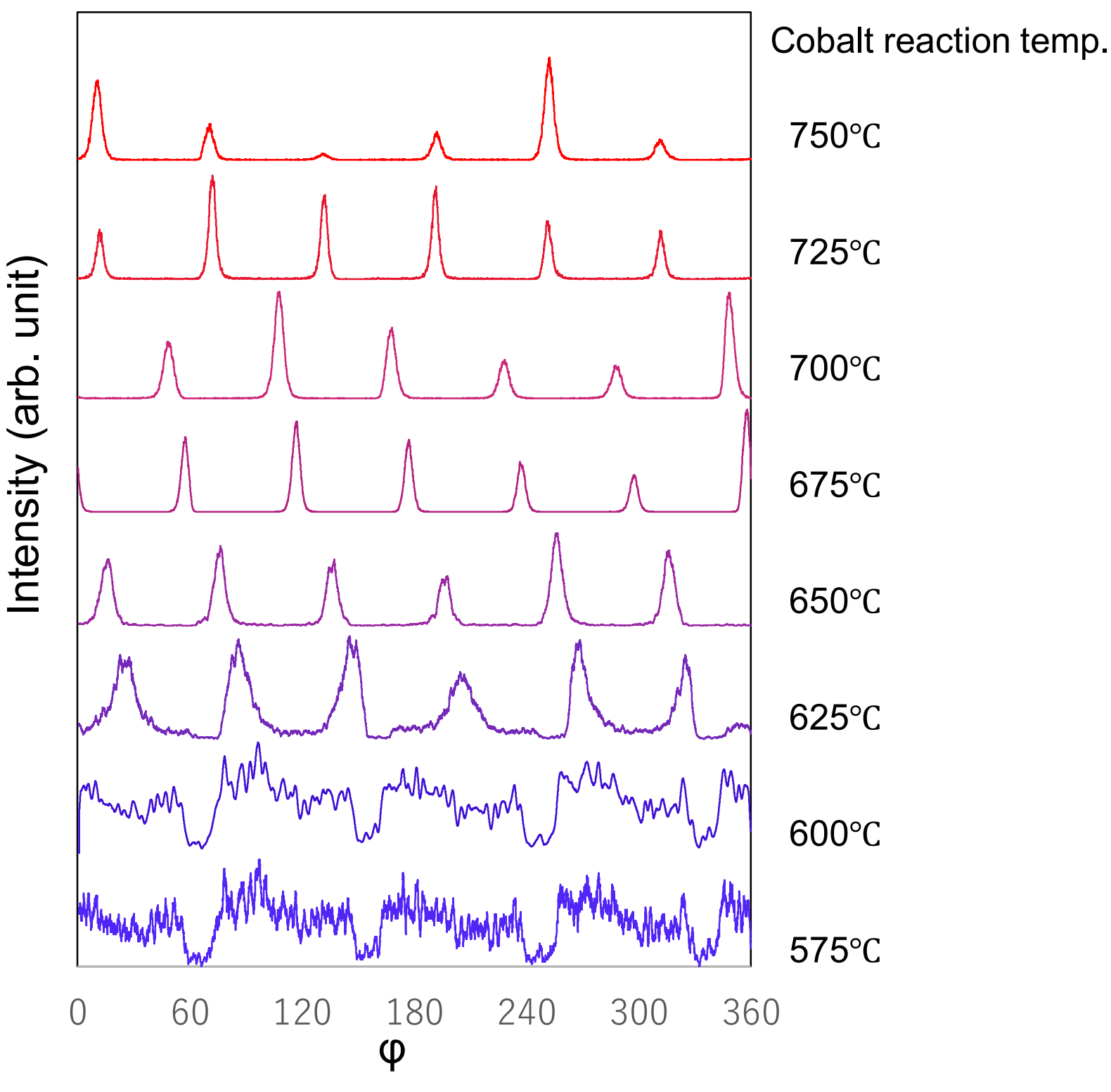

**Figure 2. Effect of annealing temperature on GaN epitaxy.** XRD φ-scans of GaN films grown by sputter deposition on Si(111) substrates. A 0.5-nm Co layer was deposited on the substrate and subsequently annealed at the indicated temperatures to form a template prior to GaN growth.

**Fig. 2** presents the XRD φ-scan results, which clearly reveal that GaN epitaxy is critically dependent on the annealing temperature. For samples annealed at or below 600°C, no six-fold symmetry was observed. This demonstrates that the as-deposited metal film itself is inert; rather, it is the cobalt silicide layer [47–51], formed during high-temperature annealing, that facilitates the epitaxial growth of GaN.

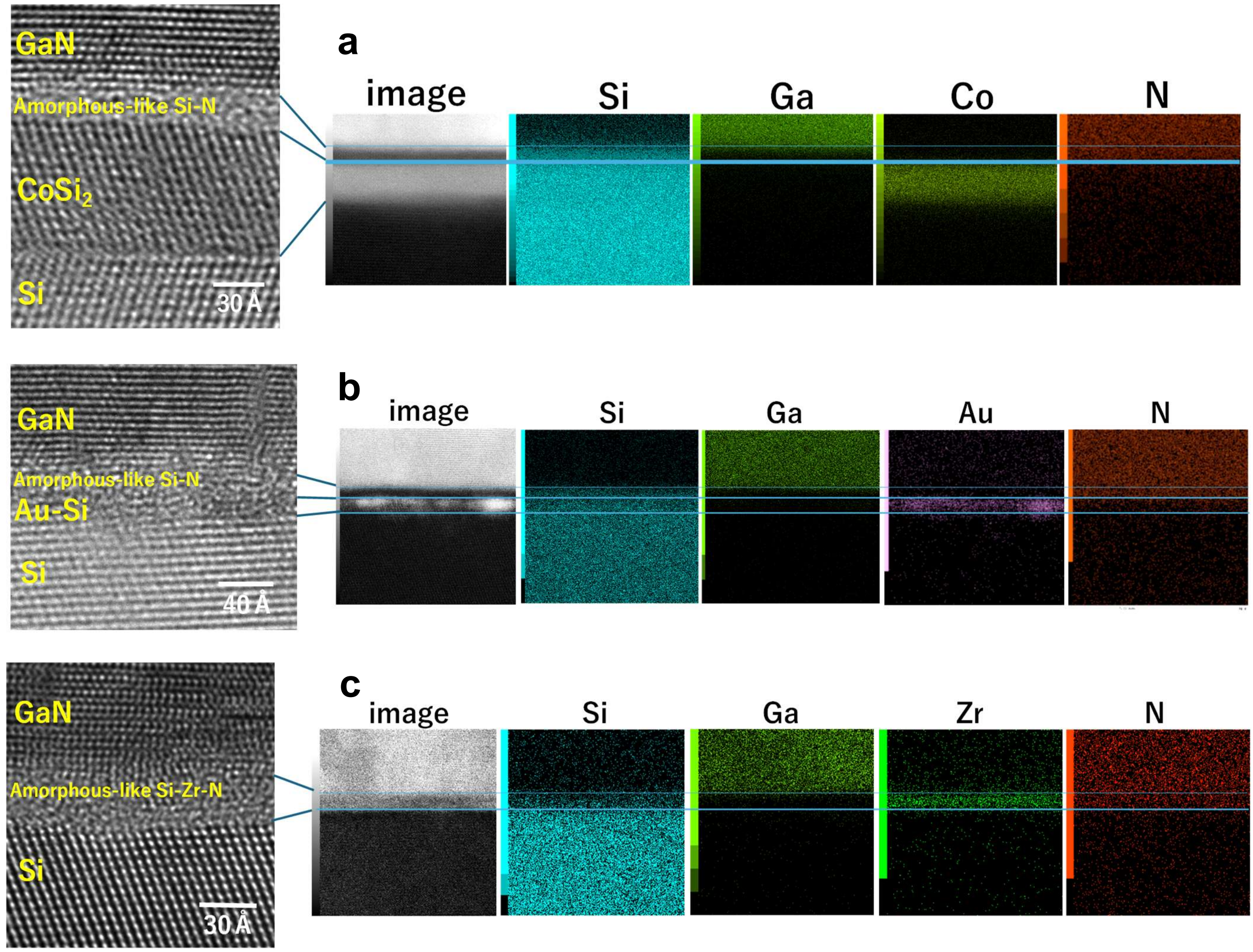

**Figure 3. STEM image and EDS mapping analysis of the epitaxial GaN/buffer/Si(111) interface.** We analyzed the interfaces of sputtered GaN films. The underlying templates were prepared by depositing a) 7-nm Co, b) 0.5-nm Au, and c) 0.5-nm Zr, which were then annealed in vacuum at 725 °C for 10 min prior to GaN growth.

We investigated the GaN/buffer/Si(111) interfacial structure using scanning transmission electron microscopy (STEM). **Fig. 3** presents cross-sectional STEM images for representative templates formed with Co, Au, and Zr, along with their corresponding energy-dispersive X-ray spectroscopy (EDS) elemental maps. Intriguingly, a common feature is observed across all samples: the formation of a distinct amorphous-like interlayer (AL-IL) directly beneath the GaN film. However, the composition of this AL-IL varies with the initial metal used: for Co and Au templates, it is primarily composed of Si and N, whereas for the Zr template, it consists of Si, Zr, and N. Beneath this AL-IL, the underlying structure reflects the specific metal-Si interaction; the Co-template exhibits a distinct crystalline $CoSi_2$ phase, whereas the Au-template shows a non-stoichiometric Au-Si layer, consistent with the fact that Au reacts with Si without forming stable stoichiometric silicide phases [52]. Similarly, for the Zr template, although Zr itself forms silicides or nitrides depending on phase stability [53–55], the incorporation of Si and N into the Zr-containing layer promotes amorphization by interrupting the crystalline lattice formation. This results in a compositionally and functionally similar amorphous-like structure to those formed with Co and Au. Despite these varied initial interfacial reactions, the consistent formation of a Si- and N-rich AL-IL at the immediate GaN interface emerges as a unifying feature.

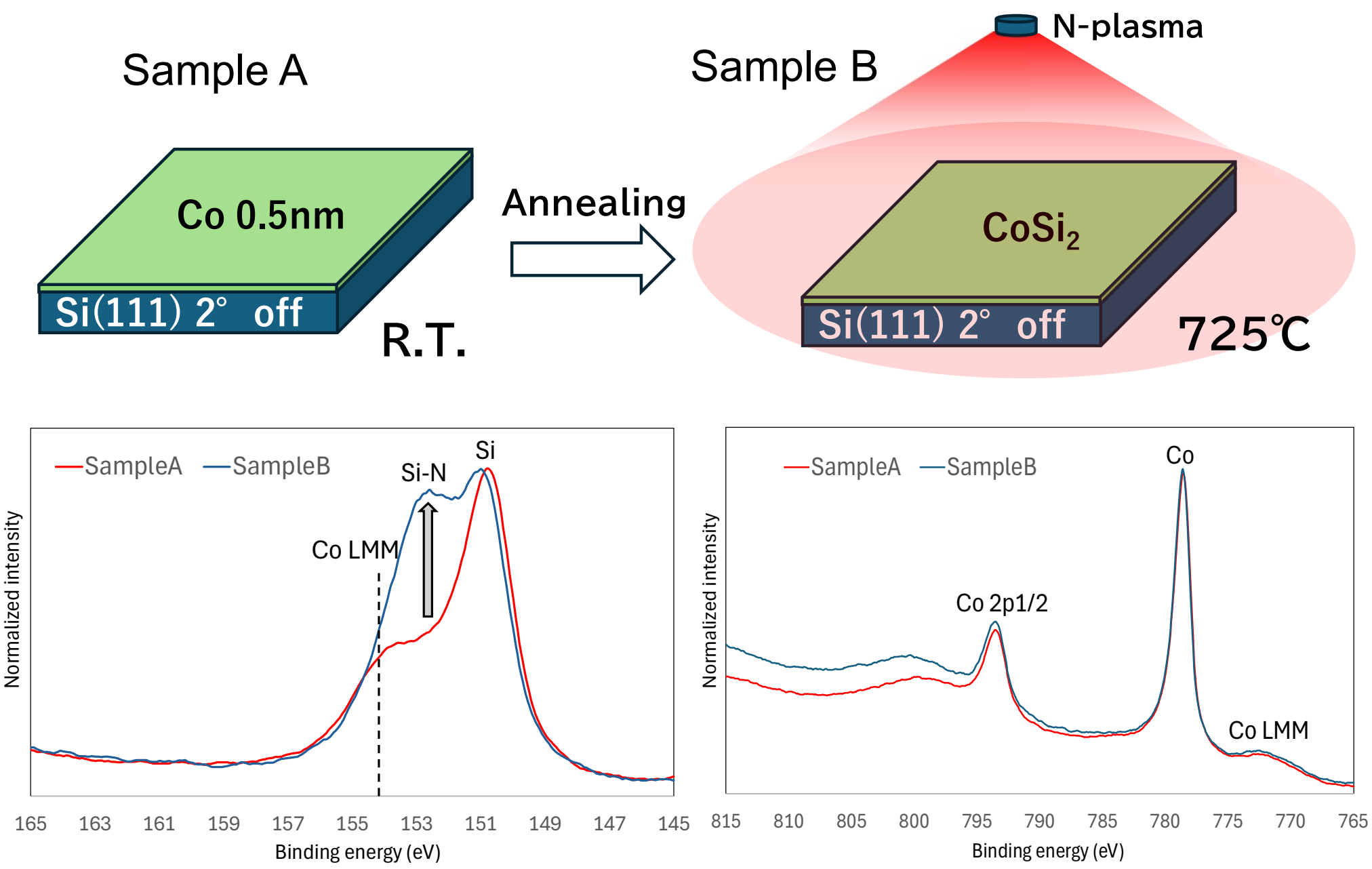

**Figure 4. XPS spectra obtained after annealing and nitrogen plasma irradiation of an ultrathin Co film deposited on Si(111).** A 0.5 nm-thick Co layer was deposited on a 3° off-axis Si(111) wafer, followed by nitrogen plasma irradiation during annealing at 725 °C under vacuum. The method of nitrogen plasma irradiation is described in Ref. [58].

XPS analysis provides definitive proof of the formation mechanism. As shown in **Fig. 4**, exposing a 0.5-nm Co/Si(111) sample to a nitrogen plasma for just 10 seconds at 725°C induces a clear shift in the Si 2s peak, indicative of Si–N bond formation [56], while the Co 2p peak remains completely unaffected [57]. This confirms that the formation of the Si–N-rich interlayer is an extremely rapid process that occurs via direct nitridation of the silicon (or silicide surface) as soon as GaN growth begins [58]. Specifically, it is inferred that the mechanism behind the formation of the amorphous-like interlayer (AL-IL) involves the simultaneous nitridation and amorphization of the topmost ultrathin silicide layer upon exposure to nitrogen plasma at the very early stages of GaN growth. In this process, the resulting surface becomes thermodynamically stable as either a Si–N phase (in the cases of Co and Au) or a Metal–Si–N phase (in the case of Zr).

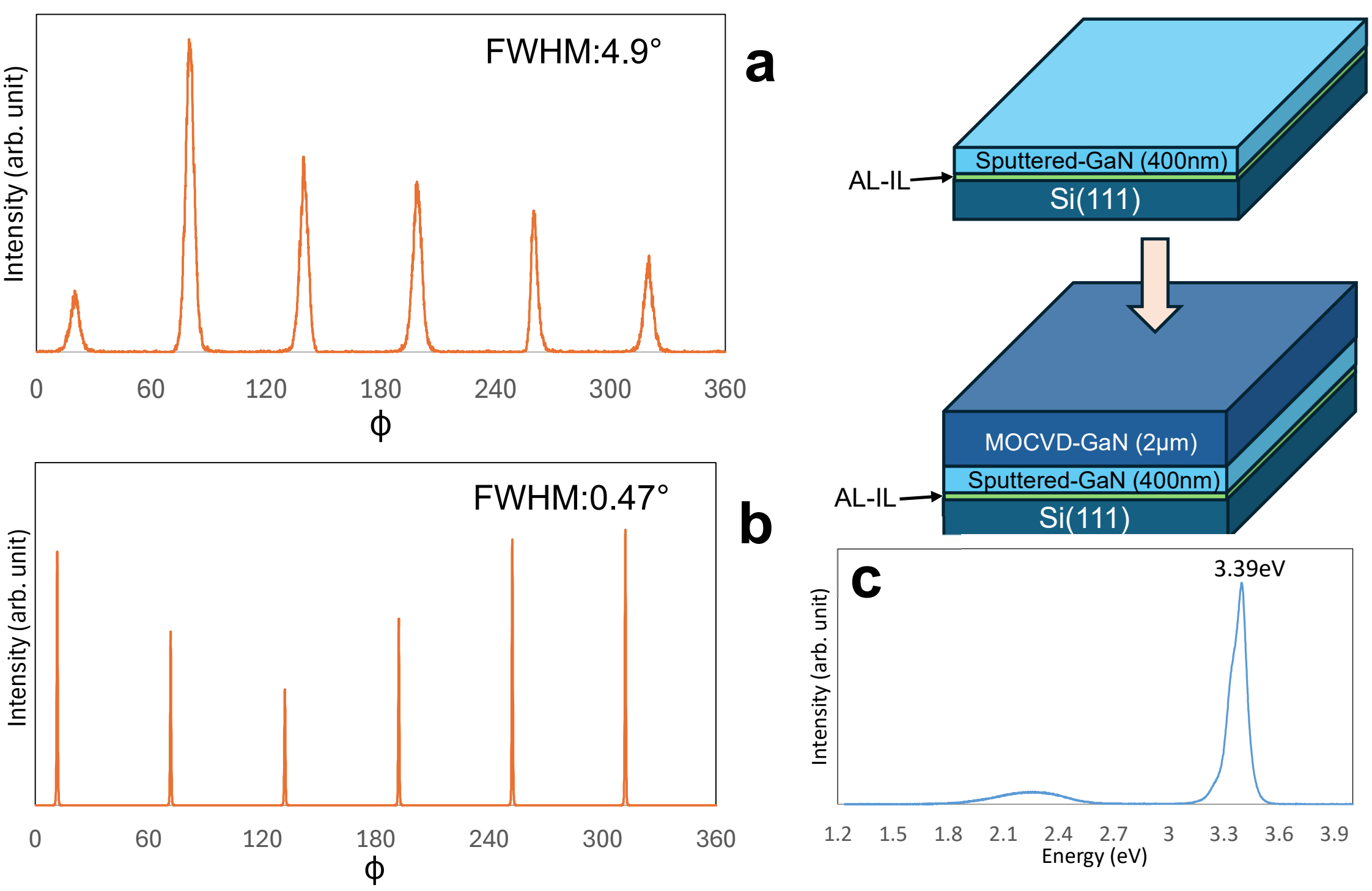

**Figure 5. MOCVD overgrowth of GaN on a sputtered GaN-on-Si(111) template.** XRD φ-scans and corresponding schematic architectures for (a) a sputtered GaN-on-Si(111) template and (b) an MOCVD-overgrown GaN film. The template was formed via a 0.5-nm Co layer, which creates an amorphous-like interlayer (AL-IL). (c) Room-temperature photoluminescence (PL) spectrum for the sample shown in (b).

Device-grade GaN and vertical conductivity **Fig. 5a-c** presents the XRD φ-scan results for a sputtered GaN film on Si(111) formed with an AL-IL, and for a 2-µm-thick GaN film subsequently overgrown on this template by MOCVD. The data reveal a significant improvement in the in-plane crystalline alignment after MOCVD overgrowth, with a full width at half maximum (FWHM) of 0.47°. Photoluminescence spectroscopy (Fig. 5c) confirms the film's good crystalline quality, as evidenced by a strong near-band-edge (NBE) emission [59–64].

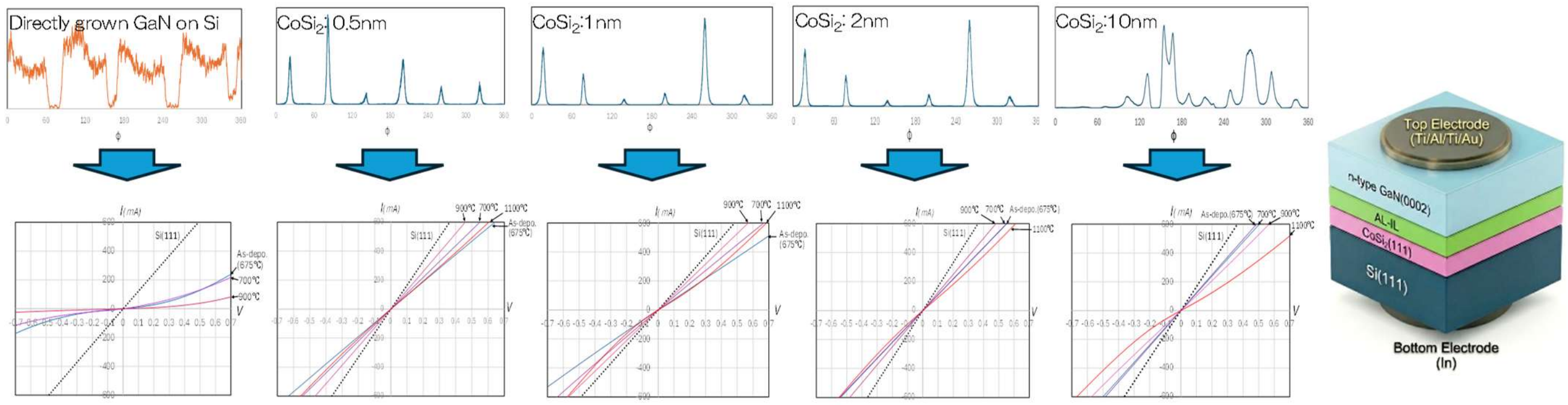

**Figure 6. Vertical current–voltage characteristics of GaN on Si(111) grown with an amorphous-like interlayer.** The plot shows vertical *I–V* curves as a function of post-growth annealing temperature. The GaN film was grown at 675 °C on a Si(111) substrate, which was pre-deposited with and without a Co layer.

**Fig. 6** presents the vertical current-voltage (*I-V*) characteristics of epitaxially grown GaN films on Si(111) by sputtering. Notably, the sample prepared with an ultrathin Co layer exhibits excellent ohmic behaviour. Furthermore, the resistance under vertical bias progressively decreases as the sample is annealed at higher temperatures. The achievement of ohmic contact through our AL-IL technique, coupled with the reduction in resistance while maintaining ohmic characteristics even after high-temperature annealing, represents a highly promising step toward the realization of vertical GaN-on-Si power devices. Moreover, the fact that a low-resistance, ohmic interface is formed despite the AL-IL being an intrinsically high-resistivity Si-N-based layer suggests that tunneling current is the dominant transport mechanism. We attribute this behavior to the in-situ Si doping of the GaN layer; specifically, the Si-rich AL-IL in contact with the GaN likely promotes the diffusion of Si atoms into the GaN interface, significantly increasing the local carrier concentration and narrowing the depletion region to facilitate tunneling. However, to further validate this mechanism and rule out other potential conduction paths, such as nanoscale pinholes or compositional inhomogeneities within the interlayer, more detailed electrical characterization—including temperature-dependent *I-V* measurements—will be conducted in future studies.

## 3. Conclusion

Here we establish a novel technique, which we term the amorphous-like interlayer (AL-IL) approach, for the direct epitaxial growth of GaN on Si(111) substrates by sputtering. Remarkably, GaN films grown using this method are not only epitaxial but also demonstrate exceptionally low vertical resistance and clear ohmic behaviour. Our findings reveal that the AL-IL formation is driven by a unique process, wherein the pre-deposition of an ultrathin metal layer onto the Si(111) surface triggers the in-situ formation of a Si-N-rich amorphous-like layer at the onset of high-temperature GaN growth. Strikingly, this effect is not limited to a specific metal; we discovered that the formation of the AL-IL and the resulting GaN epitaxy can be achieved with the pre-deposition of any elemental metal having a melting point below 1900°C. These findings firmly establish the AL-IL approach as a powerful and versatile new platform, poised to accelerate the realization of vertical GaN-on-Si devices.

## 4. Experimental Section

### 4.1 Sputter-deposition of GaN films

GaN films were deposited to a thickness of 400 nm on ultra-low resistivity n-type Si(111) wafers with a 3° off-cut angle using an RF magnetron sputtering system. A 1-inch GaN target was sputtered at an RF power of 20 W in a mixed $Ar/N_2$ (1:1 flow ratio) at a total pressure of 0.8 Pa.

### 4.2 Deposition of ultrathin metal layers and AL-IL formation

Prior to GaN deposition, ultrathin metal layers were deposited using their respective metal targets. The deposition rate for each metal was pre-calibrated to achieve a nominal thickness of 0.5 nm. The AL-IL was formed in-situ through a specific thermal sequence. The substrate with the metal layer was rapidly heated to the target temperature within 2 minutes and held for 10 minutes. GaN deposition was subsequently initiated, during which the AL-IL formed automatically. Nitrogen atoms constituting the AL-IL were supplied only at the onset of GaN deposition.

### 4.3 MOCVD growth

For MOCVD growth, a 60-nm-thick low-temperature (LT) GaN buffer layer was first deposited on the sputtered GaN template. The temperature was then increased to 1170 °C to grow a 2-μm-thick GaN layer using trimethylgallium (TMGa) and ammonia ($NH_3$).

### 4.4 Characterization

The in-plane alignment was evaluated by XRD φ-scans using the asymmetric (10-11) reflection (PANalytical X'Pert Pro MRD). Cross-sectional STEM and EDS were performed using a JEOL JEM-ARM200F operated at 200 kV. XPS analysis was performed using a ULVAC-PHI Quantera-SXM system.

PL measurements used a He-Cd laser (325 nm).

**Data availability**

Data supporting the results in this paper and the Supplementary Information are available on request to the corresponding author. Source data are provided with this paper.

**References**

[1] H. Amano, Y. Baines, E. Beam, et al., J. Phys. D: Appl. Phys. **2018**, *51*, 163001.

[2] E. A. Jones, F. F. Wang, D. Costinett, IEEE J. Emerg. Sel. Top. Power Electron. **2016**, *4*, 707.

[3] T. Oka, T. Ina, Y. Ueno, J. Nishii, Appl. Phys. Express **2015**, *8*, 054101.

[4] C. Langpoklakpam, A. C. Liu, Y. K. Hsiao, C. H. Lin, H. C. Kuo, *Micromachines* **2023**, *14*, 1937.

[5] S. X. Jin, J. Li, J. Y. Lin, H. X. Jiang, Appl. Phys. Lett. **2000**, *77*, 3236.

[6] P. Tian, J. J. D. McKendry, Z. Gonget al., Appl. Phys. Lett. **2012**, *101*, 231110.

[7] J. Yu, F. Xu, T. Tao, et al., *IEEE Electron Device Lett.* **2023**, *44*, 281.

[8] S. K. Sahu, K. Mazumdar, *J. Electron. Mater.* **2025**, *54*, 10343.

[9] V. Leitgeb, L. Mitterhuber, B. K. Legensteinet al., *J. Appl. Phys.* **2025**, *138*, 075104.

[10] P. H. Than, T. Q. Than, Y. Takaki, Mater. Adv. 2025, 6, 3139.

[11] M. Bockowski, M Iwinska, M Amilusik et al., Semicond. Sci. Technol. **2016**, *31*, 093002.

[12] Y. Tian, Y. Shao, Y. Wu et al., Sci. Rep. **2015**, *5*, 10748.

[13] Y. Oshima, T. Eri, M. Shibata, H. Sunakawa, K. Kobayashi, Jpn. J. Appl. Phys. **2003**, *42*, L1.

[14] A. Usui, H. Sunakawa, A. Sakai, A. A. Yamaguchi, Jpn. J. Appl. Phys. **1997**, *36*, L899.

[15] Y. Lai, D. Wang, Q. Kong, X. Wang, T. J. Baker, *J. Cryst. Growth* **2021**, *573*, 126216.

[16] R. Dwiliński, R. Doradziński, J. Garczyński, et al., J. Cryst. Growth **2008**, *310*, 3911.

[17] K. Grabianska, R. Kucharski, M. Amilusik, M. Bockowski, J. Cryst. Growth **2024**, *647*, 127864.

[18] A. Yoshikawa, E. Ohshima, T. Fukuda, H. Tsuji, K. Oshima, J. Cryst. Growth **2004**, *260*, 67.

[19] M. Bockowski, M Iwinska, M Amilusik, et al., Semicond. Sci. Technol. **2016**, *31*, 093002.

[20] K. Xie, T. Li, G. Ren, H. Zhou, K. Xu, *J. Alloys Compd.* **2024**, *1008*, 176776.

[21] M. Zajac, P. Kaminski, R. Kozlowski et al., *Materials* **2024**, *17*, 1160.

[22] H. Yamane, M. Shimada, T. Sekiguchi, F. J. DiSalvo, J. Cryst. Growth **1998**, *186*, 8.

[23] Y. Song, F. Kawamura, T. Taniguchi, et al., Cryst. Res. Technol. **2020**, *55*, 2000042.

[24] F. Kawamura, M. Morishita, M. Tanpo,et al., J. Cryst. Growth **2008**, *310*, 3946.

[25] M. Imanishi, et al., Cryst. Growth Des. **2025**, *25*, 6277.

[26] F. Kawamura, et al., J. Mater. Sci.: Mater. Electron. **2005**, *16*, 29.

[27] G. Huang, C. Yang, R. Pan, et al., *J. Alloys Compd.* **2025**, *1027*, 180582.

[28] M. Imanishi, S. Usami, K. Murakami, et al., *Phys. Status Solidi RRL* **2024**, *18*, 2400106.

[29] F. Kawamura, M. Morishita, N. Miyoshi, et al., J. Cryst. Growth 2009, 311, 4647.

[30] E. M. Chumbes, A. T. Schremer, J. A. Smart, et al., IEEE Trans. Electron Devices **2001**, *48*, 420.

[31] R. Stoklas, D. Gregušová, J. Novák, A. Vescan, P. Kordoš, Appl. Phys. Lett. **2008**, *93*, 124103.

[32] A. Essaoudi, H. Mosbahi, A. Gassoumi, F. Jabli, M. S. Al-Fakeh, M. Gassoumi, *Semiconductors* **2024**, *58*, 972.

[33] Y. J. Choi, S. P. R. Mallem, Y. N. Lee, K.-S. Im, S. J. An, *Surfaces Interfaces* **2026**, *80*, 108287.

[34] H. Hahn, G. Lükens, N. Ketteniss, H. Kalisch, A. Vescan, Appl. Phys. Express **2011**, *4*, 114102.

[35] H. Li, Q. Xie, Z. Lu, Y. Zheng et al., *IEEE Electron Device Lett.* **2025**, *46*, 1749.

[36] H. Qian, Y. Sun, Q. He et al., *Appl. Phys. Lett.* **2025**, *127*, 253502.

[37] Y. Yan, J. Huang, L. Pan et al., *Inorganics* **2024**, *12*, 207.

[38] A. S. Razeen, G. Yuan, J. Ong et al., *Vacuum* **2024**, *219*, 112704.

[39] P. Nautiyal, P. Pande, V. S. Kundu et al., *Microelectron. Reliab.* **2022**, *139*, 114800.

[40] Y. Zhang, M. Yuan, N. Chowdhury, K. Cheng, T. Palacios, *IEEE Electron Device Lett.* **2018**, *39*, 715.

[41] S. Michler, Y. Hamdaoui, S. Thapa et al., Phys. Status Solidi A **2025**, *222*, 2400544.

[42] R. Armitage, Q. Yang; H. Feick al., Appl. Phys. Lett. **2002**, *81*, 1450.

[43] Y. Ma, H. Chen, S. Zhan et al., Appl. Phys. Lett. **2025**, *127*, 052108.

[44] R. Zhu, H. Jiang, C. W. Tang et al., *Appl. Phys. Express* **2022**, *15*, 121004.

[45] K. Mukherjee, M. Borga, M. Ruzzarin et al., *Appl. Phys. Express* **2020**, *13*, 024004.

[46] D. Biswas, N. Torii, K. Yamamoto et al., *Semicond. Sci. Technol.* **2019**, *34*, 095013.

[47] G. J. Van Gurp, C. Langereis, J. Appl. Phys. **1975**, *46*, 4301.

[48] A. E. Morgan, E. K. Broadbent, M. Delfino et al., J. Electrochem. Soc. **1987**, *134*, 925.

[49] J. Y. Shim, S. W. Park, H. K. Baik, *Thin Solid Films* **1997**, *292*, 31.

[50] L. J. Chen (Ed.), *Silicide Technology for Integrated Circuits*, IET, **2004**.

[51] S. P. Murarka, *Silicides for VLSI Applications*, Academic Press, **1983**.

[52] W. Klement, R. H. Willens, P. Duwez, *Nature* **1960**, *187*, 869.

[53] A. V. Tkachenko, T. Ya. Kosolapova, Sov. Powder Metall. Met. Ceram. **1968**, *7*, 178.

[54] P. J. Martin, A. Bendavid, J. M. Cairney, M. Hoffman, Surf. Coat. Technol. **2005**, *200*, 2228.

[55] J. Musil, R. Daniel, P. Zeman, O. Takai, *Thin Solid Films* **2005**, *478*, 238.

[56] G. M. Ingo, N. Zacchetti, D. della Sala, C. Coluzza, J. Vac. Sci. Technol. A **1989**, *7*, 3048.

[57] E. C. Mattson, D. J. Michalak, W. Cabrera, J. F. Veyan, Y. J. Chabal, Appl. Phys. Lett. **2016**, *109*, 091602.

[58] Y. Song, F. Kawamura, K. Shimamura, T. Ohgaki, N. Ohashi, *AIP Adv.* **2020**, *10*, 115011.

[59] T. Onuma, N. Sakai, T. Okuhata, A. A. Yamaguchi, T. Honda, Phys. Status Solidi C **2011**, *8*, 2321.

[60] K. Omae, et al., Jpn. J. Appl. Phys. **2004**, *43*, L173.

[61] Y. Kawakami, et al., Appl. Phys. Lett. **1996**, *69*, 1414.

[62] H. Amano, N. Sawaki, I. Akasaki et al., *Appl. Phys. Lett.* **1986**, *48*, 353.

[63] W. Wang, H. Yang, G. Li, *J. Mater. Chem. C* **2013**, *1*, 4070.

[64] T. J. Badcock, M. Ali, T. Zhu et al., *Appl. Phys. Lett.* **2016**, *109*, 151110.

**Funding**

This study was supported by Innovative Science and Technology Initiative for Security (Grant

Number JPJ004596), ATLA, Japan.

This work was supported by Adaptable and Seamless Technology transfer Program through Target-driven R&D (A-STEP) from Japan Science and Technology Agency (JST) Japan, Grant Number JPMJTR25T5.

A part of this work was supported by "Advanced Research Infrastructure for Materials and Nanotechnology in Japan (ARIM)" of the Ministry of Education, Culture, Sports, Science and Technology (MEXT). Proposal Number JPMXP1225NM5364.

**Acknowledgements**

K.M would like to thank Ms. M. Taketomi and Ms. Y. Nakayama for their support with TEM sample preparation

**Author contributions**

F.K. conceived the project, synthesized the thin films and performed XPS, XRD and electrical characterization. K.M. conducted STEM analysis and the associated EDS elemental analysis. T.O. performed photoluminescence measurements.

**Competing interests**

The authors declare no competing interests.